\documentclass{aa}  
\usepackage{graphicx}
\usepackage{natbib}
\bibpunct{(}{)}{;}{a}{}{,} 

\usepackage{txfonts}
\usepackage[]{hyperref}
\usepackage{epstopdf}
\usepackage{todonotes}
\usepackage{multirow}

\begin{document} 

   \title{Combining FOF and halo-based algorithms for the identification of galaxy groups}
   \titlerunning{Combining FOF and halo-based algorithms}

   \author{Facundo Rodriguez \thanks{facundo.rodriguez@unc.edu.ar}\and Manuel Merchán}
   
   \authorrunning{F. Rodriguez \& M. Merchán}
   
\institute{ Universidad Nacional de Córdoba. Observatorio Astronómico de Córdoba. Córdoba, Argentina
        \\
  CONICET. Instituto de Astronomía Teórica y Experimental. Laprida 854, X5000BGR, C\'ordoba, Argentina.
          }
             
   \date{\today}

\abstract
{Galaxy groups provide the means for a great diversity of studies that contribute to a better understanding of the structure of the universe on a large scale and allow the properties of galaxies to be linked to those of the host halos. However, the identification of galaxy systems is a challenging task and therefore it is necessary to improve the techniques involved as much as possible.}
{In view of the large present and forthcoming galaxy catalogues, we propose, implement, and evaluate an algorithm that combines the two most popular techniques used to identify galaxy systems. The algorithm can be easily applied to any spectroscopic galaxy catalogue, but here we demonstrate its use on the Sloan Digital Sky Survey.}
{Assuming that a galaxy group is a gravitationally bounded system that has at least one bright galaxy, we begin by identifying groups with a Friends-Of-Friends algorithm adapted to fit this definition. In view of the shortcomings of this method, particularly the lack of ability to identify low-number groups, and consequently the inability to study the occupation of halos throughout the mass range, we improve it by adding a halo-based procedure. To assess the performance, we construct a mock catalogue from a semi-analytical model to compare the groups identified using our method with those obtained from the simulation.} 
{The comparison of  groups extracted using our method with those of a mock catalogue shows that the proposed algorithm provides excellent results. The modifications introduced to the Friends-Of-Friends algorithm in the first part of the procedure to fit the adopted group definition gave reliable groups. Furthermore, incorporation of the halo-based method reduces the interlopers while more accurately reproducing the number of galaxies per group. As a specific application, we use the algorithm to extract groups from the Sloan Digital Sky Survey.} 
{}


   \maketitle
%
\section{Introduction}

The current paradigm assumes that galaxies form by baryon condensation within the potential wells defined by the collisionless collapse of dark matter halos \citep{White1978}. Studies of galaxy groups allow  galaxies to be linked to the halos in which they reside. Therefore, improving the identification of galaxy groups will enhance our ability to study the large-scale structure of the universe. Because hierarchical clustering induces structure formation, many studies require reliable group information. Some examples are lens analysis \citep{Jullo2010,Leauthaud2017,Gonzalez2019}, and studies of galaxy formation and evolution \citep{Kawata2007,Muriel2014,Kanagusuku2016, Taverna2016}, and the large structures of the universe \citep{Paz2011,Luparello2013,Ceccarelli2013,Sutter2014, Martinez2015}. 

Several studies have shown that due to the hierarchical assembly of halos and galaxy mergers, the groups generally present a central or main galaxy with more stellar mass and/or luminosity surrounded by other satellite galaxies \citep{Larson1980,Balogh2000, Mccarthy2007,Campbell2015,Hearin2017}. A powerful indicator used to study the central and satellites galaxy populations in dark matter halos is the halo occupation distribution \citep[HOD,][]{Berlind2002,Kravtsov2004, Zehavi2005, Zheng2005,Rodriguez2015,Zehavi2018,Vakili2019}, and in particular the halo occupation function which represents the average number of galaxies for a given halo mass, $\langle$N|M$\rangle$. However, to carry out direct HOD determinations across the whole mass range in which the halos contain galaxies, poor and numerous reliable galaxy groups are needed. In addition, improving the identification of galaxy groups allows us to populate simulations that more accurately adjust the way galaxies inhabit dark matter halos using the correlation function and the HOD \citep[an example of this is presented in ][]{Carretero2014}. 

There are several methods to extract groups of galaxy catalogues: Friends-Of-Friends \citep[FOF; ][]{Huchra1982}, Halo-based \citep{Yang2005}, Voronoi-Delaunay-based \citep{Gerke2012, Pereira2017}, matched filter \citep{Kepner1999,Milkeraitis2010}, density field-based \citep{Miller2005,Sharma2009, Smith2012}, and Bayesian-based using the marked point process framework \citep{Tempel2018}, among others. In this work, we focus our attention on improving the redshift space identification of groups obtained by the two most used techniques at this time: FOF and Halo-based.

The FOF method proposed by \cite{Huchra1982} is based on spatial criteria for assigning galaxies to groups. Different versions of this technique
have been widely used and applied to several catalogues; for example by \citet{Zeldovich1982}, \citet{Maia1989}, \citet{Ramella1989}, \citet{Ramella1997}, \citet{Ramella2002}, \citet{Nolthenius1987}, \citet{Trasarti1998}, \citet{Giuricin2000}, 
\citet{Adami2002}, \citet{Merchan2002}, \citet{Eke2004}, \citet{Merchan2005}, \citet{Berlind2006}, \citet{Tago2006}, \citet{Deng2007}, \citet{Tago2008}, \citet{Tago2010}, \citet{Calvi2011}, \citet{Munoz2012}, \citet{Tempel2012}, \citet{Tempel2014}, \citet{Tempel2017}, among others. As this method is probably the most commonly used to identify groups of galaxies, several authors have focused on the analysis of its performance, such as for example \citet{Frederic1994} and, more recently, \citet{Nurmi2013}, \citet{Duarte2014}, \citet{Old2014}, \citet{Old2015} ,\citet{Old2018}, \citet{Wojtak2018}, \citet{Davies2019}. Beyond the demonstrated virtues of this algorithm and its flexibility to adapt to different scientific objectives (e.g., transverse and perpendicular linking lengths, scale factor), one of the negative aspects that can be extracted from these analyses is that the poor groups  (that have 4 galaxies or less)
 obtained with this algorithm have low reliability and those which are numerous (more than 10 galaxies) are likely to have many interlopers. An attempt to improve FOF performance was presented in \citet{Tempel2014,Tempel2016}, which starts from a conventional FOF followed by the application of a double refinement, first applying a multimodal analysis based on the distribution of galaxies in the phase space and then an evaluation of the membership based on the virial radius and escape velocity. Even with these types of improvements, these catalogues are not suitable for certain studies, such as for example HOD analysis.
 
The halo-based group finder developed by \cite{Yang2005} and applied to the SDSS \citep{Yang2007, Duarte2015}, has been increasingly used in recent years because it finds groups with few members. In this method, the luminosity of the galaxies is related to the mass of the dark matter halos, assigning masses using the abundance-matching technique on luminosity. The group catalogues from these identifications allow the study of the HOD in a wide range of masses and luminosities. However,  the method starts with a FOF algorithm with an overdensity that is not representative of the galaxy groups and an imposed mass--luminosity ratio, and this can bias the characteristics of the resulting group catalogue. Other authors who have similarly implemented this method recently are Duarte and Mamón (2015), who do not start from a FOF but instead take bright or high-stellar-mass galaxies as potential group centres to start the iteration.

Here, we propose a method to identify galaxy groups that combines the two methods described above: a proper FOF identification is performed in the first step and, using the resulting groups, in the second step we implement an algorithm that follows the halo-based method. Understanding each step as an improvement of the previous one, we asses the performance of this process applying this algorithm to a mock catalogue. This procedure allows us to learn about the reliability of the obtained groups. Finally, taking into account this information, we apply this algorithm to the Sloan
Digital Sky Survey Data release 12 \citep[SDSS DR12;][]{Alam2015}, in particular to the spectroscopic galaxy catalogue. 

Our work is organized as follows: In Section \ref{sec:desc}, 
we describe both parts of our identification algorithm.
In Section  \ref{sec:mock}, we asses the performance of our proposed group finder. 
In Section \ref{sec:real}, we describe the implementation of our algorithms in the SDSS DR12 and we compare some of our results with those obtained by \cite{Tempel2017} and \cite{Yang2008}. Finally, in section \ref{sec:conc} we present a summary and our conclusions.

\section{Proposed group finder}
\label{sec:desc}

The proposed algorithm consists of two parts: first, we apply a FOF algorithm with some modifications and then, starting from these groups, we implement a halo-based group finder.
The method is presented in this way because it is easier to explain, but also because we use these stages below to evaluate the benefits of each part of the procedure.

\subsection{Part I: FOF implementation}
We began our group identification using the FOF algorithm implemented by \cite{Merchan2002, Merchan2005}. Given a pair of galaxies with mean radial comoving distance $D=(D_1 + D_2)/2$ and angular separation $\theta_{12}$, the algorithm links galaxies satisfying the conditions:
\begin{equation}
D_{12}=2\sin\left( \frac{\theta _{12}}{2}\right) D \leq D_L=D_0 R,
\label{eq:fof-1}
\end{equation}
and
\begin{equation} \label{eq:fof-2}
V_{12}=|V_1-V_2|\leq V_L=V_0 R,
\end{equation}
where $D_{12}$ is the projected distance and $V_{12}$ is the line-of-sight
velocity difference. As can be seen in the above formulae, taking into account the number density variation due to the apparent magnitude limit of the survey, the transverse ($D_L$) and radial ($V_L$) linking lengths scale with $R$. That factor is determined using the galaxy luminosity function of the sample $\phi (M)$ :
\begin{equation} \label{eq:fof-3}
R = \left[ \frac{\int_{-\infty} ^{M_{12}} \phi (M) dM}{\int_{-\infty} ^{M_{lim}} \phi (M) dM}\right]^{-\frac{1}{3}}
,\end{equation}
where $M_{lim}$ and $M_{12}$ are the absolute magnitude of the brightest
galaxy visible at the fiducial ($D_f=50\,\rm{Mpc}$) and mean galaxy ($D$) distances, respectively, and $D_0$ is chosen according to the desired overdensity:
\begin{equation} \label{eq:fof-4}
\frac{\delta \rho}{\rho} =\frac{3}{4\pi D_0^3}\left[ \int _{-\infty} ^{M_{lim}} \phi (M) dM \right]^{-1} -1
.\end{equation}
We choose the value of $ V_0 = $ 200 km s$^{-1}$ following \cite{Merchan2002}, who conclude that this value is a good compromise between the number of detected groups and the contamination by false groups. Throughout this work, we use an overdensity of  $\delta \rho /\rho =\Delta _{200} =200$.

After the implementation of this algorithm, we have a sample of galaxy groups with two or more members. That is, we have the centre of these groups, the galaxies that compose them, and the properties of each of them. However, as is well known, the groups determined following a FOF algorithm are more reliable as they have more galaxies \citep{Merchan2002, Merchan2005}. 
To improve the reliability of poor groups, we restrict the luminosity of group members: they must have at least one bright galaxy to be considered as a group.\footnote{This choice is consistent with what is pointed out in \cite{Tempel2009}: almost all bright isolated galaxies can be identified with most luminous galaxies where the remaining galaxies lie outside the observational window used in the selection of galaxies for the survey. Truly isolated galaxies are rare; they are faint and are located mainly in voids.} Throughout this work, we consider a bright galaxy as all those that have an absolute magnitude in r-band $M_r<M_{{gr}_{lim}}=-19.5$. With this in mind, we reject all groups that do not meet this criterion and, additionally, we add all bright galaxies as potential groups. With these restrictions, our group definition is not limited to simple geometric considerations but also takes into account the astrophysical properties of the galaxies that comprise it. Thus, for us, a group of galaxies is a system of galaxies gravitationally bounded that has at least one bright galaxy. 

\subsection{Part II: Halo-based group finder}

This part of the method is performed following the halo-based group finder proposed by \cite{Yang2005, Yang2007}, which is iterative and based on an adaptive filter modelled after the general properties of dark matter halos. As this method requires a catalogue of potential groups and our intention is to improve the reliability of the previously identified systems, we use the resultant galaxy groups of part I and, based on its properties, implement a similar iterative procedure.

\vspace{0.25cm}
\textit{1. Estimation of the characteristic luminosity.}

Because further below we need luminosities to assign masses and we want to be consistent with the previous group definition, we set the characteristic luminosity ($L_{gr}$) of the group  as the sum of the luminosities of all group members with a magnitude equal to or brighter than $M_{{gr}_{lim}}$. 

As there are systems that are  further away meaning that we cannot see all their galaxies, it is necessary to correct the characteristic luminosity \citep{Moore1993}. Therefore, we define $L_{gr}$ as the sum of the observed luminosity $L_{obs}$ plus a correction $L_{cor}$, 
\begin{equation} 
L_{gr}=L_{obs}+L_{cor}
,\end{equation}
where $L_{obs}$ is the sum of the luminosities of the observed galaxies and $L_{cor}$ is the expected luminosity corresponding to invisible members:
\begin{equation} 
L_{corr}= N_{cor}  \frac{\int_{L_{-19.5}} ^{L_{lim}} L \, \phi (L)\, dL}{\int_{-\infty} ^{L_{lim}} \phi (L)\, dL}
,\end{equation}
where $L_{lim}$ is the luminosity of the faintest galaxy that can be seen at the  redshift of the group; $L_{-19.5}$ is the luminosity corresponding to a magnitude $M_{{gr}_{lim}}$; and $N_{cor}$ is the number of observed galaxies.
Because in order to estimate $L_{gr}$ we only consider the brighten galaxies than $M_{{gr}_{lim}}$ and, given the flux-limited nature of galaxy catalogues, $L_{cor}$ only takes effect beyond a given redshift $z_{lim}$. In any case, as $M_{gr}$ is quite bright, $L_{cor}$, in general, is not significant.  

\vspace{0.25cm}
\textit{2. Properties of the dark matter halo derived from its luminosity.}

To obtain the halo mass ($M_h$) from luminosity, we use an abundance-matching technique on $L_{gr}$, a procedure widely used and studied by several authors \citep[e.g.][]{Vale2004,Kravtsov2004, Tasitsiomi2004, Conroy2006, Behroozi2010, Cristofari2019}. This  technique assumes a one-to-one relationship between the characteristic luminosity and the halo mass. Consequently, the mass assignment is reduced to order halos by luminosity and associate masses according to abundance prescribed by the mass function.
As pointed out by \cite{Yang2007}, it is an over-simplification, but some studies show that it is a good approximation  overall \citep[e.g.][]{Gonzalez2019}.

Using $M_h$ given by the abundance-matching procedure and the same overdensity adopted for the implementation of part I, we obtain the radius of the halo ($r_{200}$) and  the line-of-sight velocity dispersion ($\sigma$):
\begin{equation} 
\label{eq:hb-1}
r_{200}=\left( \frac{2\,M_h G}{\Delta_ {200}\,H(z_{gr})^2}\right)^{-\frac{1}{3}}
,\end{equation}
\begin{equation} \label{eq:hb-2}
\sigma=\sqrt{\frac{M_h G}{2r_{200}}}
,\end{equation}
where $G$ is the gravitational constant and $H(z_{gr})$ is the Hubble constant at redshift group centre. Some details of how these expressions are reached are figured out in Appendix \ref{appendixx}.

To conclude this step, we use the relation provided by \cite{Ludlow2016} to calculate the halo concentration, $c_{200}$, the scale radius, $r_s$ , and the average density inside the virialization radius $\rho _{200}$ as a function of $M_h$ corresponding to an Navarro–Frenk–White (NFW) profile.
 
\textit{3. Membership assignment.}

This step is critical because we are going to dismiss FOF memberships in favour of this method, which we expect will have less interloper contamination and a better mass estimation including groups with few members. 

Following \cite{Yang2005}, we define the three-dimensional density contrast in redshift space, $P_M(D,\Delta z),$ and calculate the projected distance and redshift distance ($D$ and $\Delta z$, respectively) between the galaxy and all group centres. To derive these quantities, \cite{Yang2005} assume that the distribution of galaxies in phase space follows that of the dark matter particles. Thus, $P_M(D,\Delta z)$ can be written as:
\begin{equation} \label{eq:hb-3}
P_M(D,\Delta z)= \frac{H_o}{c} \frac{\Sigma (D)}{\bar\rho} p(\Delta z)
,\end{equation}
where $c$ is the speed of light, $\Delta z =z-z_{gr}$, $\bar \rho$ is the average density of the Universe, and $\Sigma (D)$ is the projected surface density. As we adopt a spherical NFW profile:
\begin{equation} \label{eq:hb-4}
\Sigma (D)=2r_s \bar \delta _{200} \bar \rho f(D/r_s)
,\end{equation}
where
\begin{equation} \label{eq:hb-5}
f(x)=
 \begin{cases} 
      \frac{1}{x^2-1}\left\{ 1- \frac{\ln{\left[ \left(  1+ \sqrt{1+x^2} \right) /x \right] }}{\sqrt{1-x^2} }  \right \}   & \mbox{, } x < 1 \\
      \frac{1}{3}  & \mbox{, } x = 1 \\ 
      \frac{1}{x^2-1}\left( 1- \frac{\arctan{\left(\sqrt{1-x^2}\right)}}{\sqrt{1-x^2} }  \right)  & \mbox{, } x > 1, 
   \end{cases}
\end{equation}
\begin{equation} \label{eq:hb-6}
\bar \delta _{200} =\frac{\Delta _{200}}{3}\frac{c_{200}^3}{\ln{(1+c_{200})}-c_{200}/(1+c_{200})}
.\end{equation}

The redshift distribution of galaxies within the halo is described by $p(\Delta z),$ and assuming a Gaussian behaviour:
\begin{equation} \label{eq:hb-7}
p(\Delta z)= \frac{1}{\sqrt{2\pi}}\frac{c_{200}}{\sigma (1+z_{gr})}\exp{\left[\frac{-(c_{200}\Delta z)^2}{2\sigma ^2 (1+z_{gr})} \right]}
.\end{equation}

Here, $P_M(D,\Delta z)$ allows us to model the density contrast in the position of each galaxy with respect to the centres of the potential groups. Higher values of this quantity indicate a higher probability of belonging to a group.
Thus,following \cite{Yang2007}, $P_M(D,\Delta z)$ is used to assign members to groups so that all galaxies that exceed a cut-off value of $P_M(D,\Delta z)>10$ are considered as belonging to the system. If a galaxy exceeds this value for more than one group, we choose the group with the highest $P_M$.
In the situation where all the galaxies of one group could be assigned to another, both groups are merged; later we discuss the chosen cut-off value.

\textit{4. Iteration.}

Since membership depends primarily on the total luminosity of the group, and this in turn depends on the galaxies that make up the group, an iterative process is necessary. Given that the memberships are obtained in step 3, the iteration consists of a return to the first step and repetition of all the process beginning from the estimation of the total luminosity using the newly assigned members.
We stop this iteration cycle when there are no more changes in group membership, which generally takes no more than four iterations.

To start the iterative process, we take the membership of FOF groups described in Part I to estimate
the first luminosities. In the following iterations, we use the members determined using the procedure described in Part II. Since our initial input is FOF groups, and it is well-known that groups obtained in this way are more reliable as they are more numerous,  rich groups cn be expected to have only minor changes but groups with few members should be more affected. We also trust that our algorithm is able to discard interlopers and detect groups with few members, including those systems with one or two members. 

As a result of this process, we obtain two group samples. The first is obtained from part I, which consists of a suitable FOF algorithm following \cite{Huchra1982}  adapted to our group definition and hereinafter referred to as FOF modified. The second group, which results from the whole process and  introduces a halo-based technique described in part II, is called the Final Sample.

\section{Algorithm performance}
\label{sec:mock}
To test the effectiveness of the proposed group finder, we use a mock catalogue that reproduces the features of the SDSS DR12 spectroscopic catalogue \citep{Alam2015} built up from a semi-analytic model. This allows us to understand exactly how galaxies populate dark matter halos and, with this information we can evaluate the assignment made by our identifier and estimate the errors that will need to be taken into account when applying the method to the real catalogues.

\subsection{Mock catalogue}
Our mock catalogue is built using the semi-analytical galaxy formation model developed by \cite{Guo2010} applied on the Millennium Simulation \citep{Springel2005}, which offers high spatial and time resolution within a large cosmological volume. This simulation evolves more than 10 billion dark matter particles in a 500 $h^{-1}$ Mpc periodic box, using a comoving softening length of 5 $h^{-1}$ kpc. 

To construct our mock catalogue, we place the observer at the Millennium box coordinate origin and, taking into account the periodicity, we simply repeat the simulated volume as many times as is necessary to reproduce the volume of the SDSS DR12 spectroscopic catalogue. The redshifts are calculated combining the cosmological distance and the distortion produced by proper motions. From these redshifts and the absolute magnitudes provided by the semi-analytical model, it is possible to obtain the apparent magnitudes of each galaxy. To mimic the flux-limited selection of the SDSS DR12, we impose the same upper apparent magnitude threshold. In order to imitate the angular selection function of the survey, we built a mask dividing the celestial sphere into $\sim$ 7 700 000  pixels of equal area using SDSSPix software specifically designed for the SDSS geometry \citep{Swanson2008}. We compute an initial binary mask by setting to `1' all pixels with at least one galaxy inside, and `0' the remaining pixels. Given that the pixel size is much smaller than the mean inter-galaxy separation, we smooth the initial mask by averaging each pixel over its 7 $\times$ 7  adjacent neighbours. We compound the angular selection function as the set of all pixels with a value that exceeds a given threshold. This is the same procedure used by \cite{Rodriguez2015}.

Simulation, together with the semi-analytical model, not only gives us the galaxies but also provides us with their membership to a given dark matter halo. This allows us to construct a synthetic catalogue of galaxy groups with the exact membership of each galaxy to its corresponding dark matter halo. 
As we describe in the following section, this synthetic catalogue will be highly suitable for evaluation of the performance of our group finder.

\subsection{Purity and completeness}

A traditional way to assess the quality of group-finding algorithms is through the purity and completeness concepts. Group purity, $P$, quantifies the extent to which galaxies that are identified as members of a given group  actually  reside in the same dark matter halo:
\begin{equation}
    P=\frac{n_{gh}}{N_{g}}
,\end{equation}
where $N_g$ is the total number of galaxies of the identified group and $ n_{gh} $ is the number of galaxies in the identified group $g$ that reside in the halo $h$. With this definition, the value of $1-P$ indicates the percentage of interlopers. If a group shares galaxies with more than one halo, we choose the halo with the highest number of galaxies $ n_{gh} $.

 \begin{figure*}
   \centering
   \includegraphics[width=0.49\textwidth]{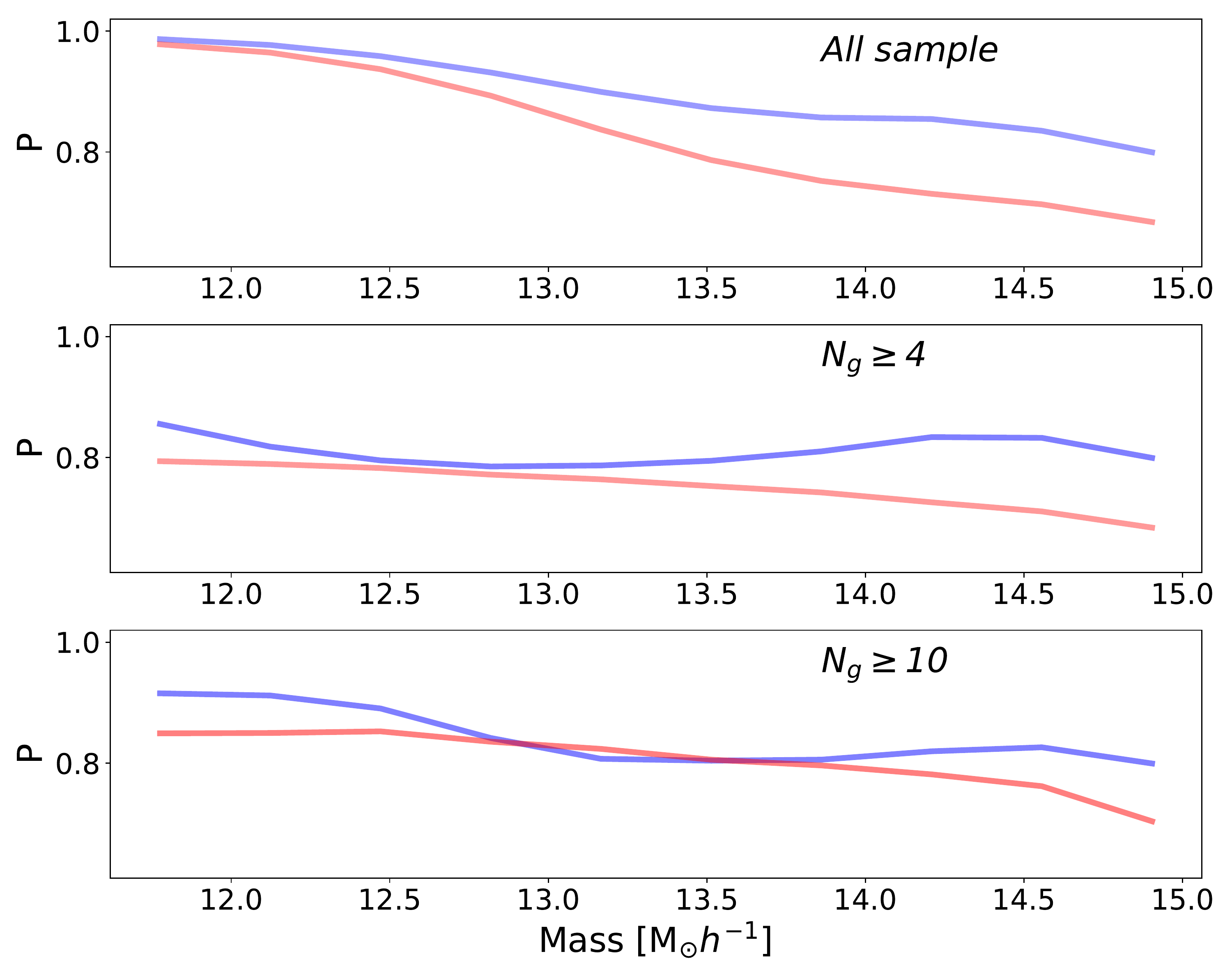}
   \includegraphics[width=0.49\textwidth]{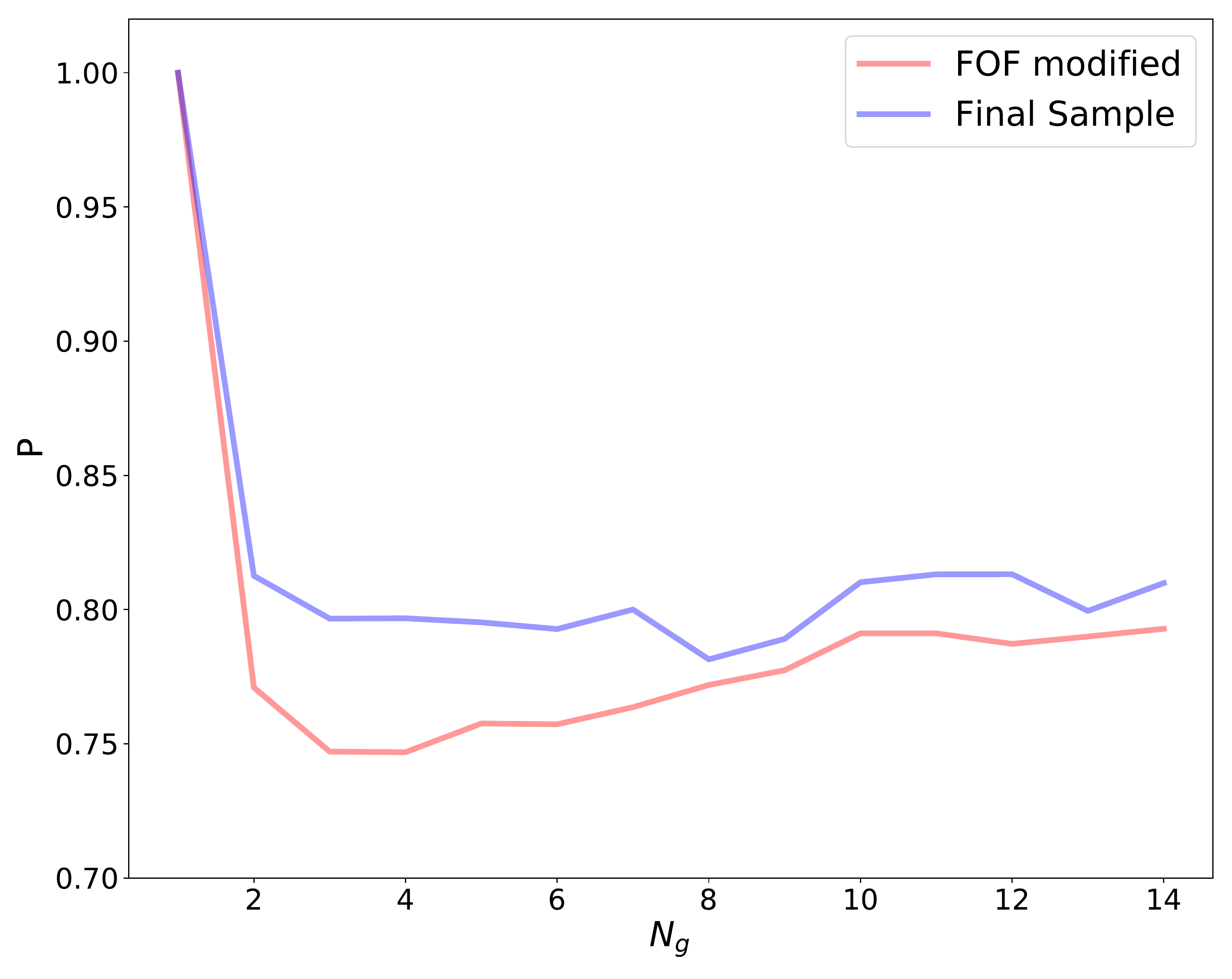}
      \caption{Group purity as a function of halo mass determined by abundance matching on luminosity for the FOF modified group catalogue (red) and Final Sample, combining FOF and halo-based methods (blue). Left panels show the results for all sample, groups with four or more members and groups with ten or more (on top, middle and bottom panels, respectively). The right panel presents group purity as a function of group multiplicity for the FOF modified group catalogue (red) and Final Sample (blue).}
         \label{Purity}
   \end{figure*}

In general, the complete identification process (part I and II of the previous section) is capable of producing high-purity groups, which is shown in the upper left panel of Figure \ref{Purity} (blue line) as a function of the mass. In addition, to show how the iterative method of part II improves identification, we include the corresponding result to apply only part I of the process (red line). As can be seen, the application of the last part of the process has a greater influence on high-mass groups. This is because these groups are more numerous and therefore have more opportunities to improve their membership. On the other hand, the opposite case occurs for low-mass groups, which are dominated by groups of one member that, given our group definition, necessarily have a purity equal to 1. To verify this statement, we also show the purity as a function of mass for groups with four or more members and ten or more (left lower panels of Figure \ref{Purity}). In both cases, the purity is approximately constant, with values similar to that of the most massive groups of the total sample. Continuing with this reasoning, it is useful to know the behaviour of purity as a function of $N_g$, which is, unlike mass, a direct result of the identification process (see right panel of Figure \ref{Purity}). Here, again, it can be seen that the groups with one member have a purity that is mandatorily 1. Beyond this consequence of our group definition, our method confers the advantage that the purity remains approximately constant for the rest of the groups. 

The next quantity to be defined is halo completeness, $C$. Conceptually, $C$ quantifies the fraction of visible galaxies in a dark matter halo that are included in some identified galaxy group:
\begin{equation}
    C=\frac{n_{hg}}{N_{h}} 
,\end{equation}

   \begin{figure*}
   \centering
   \includegraphics[width=0.49\textwidth]{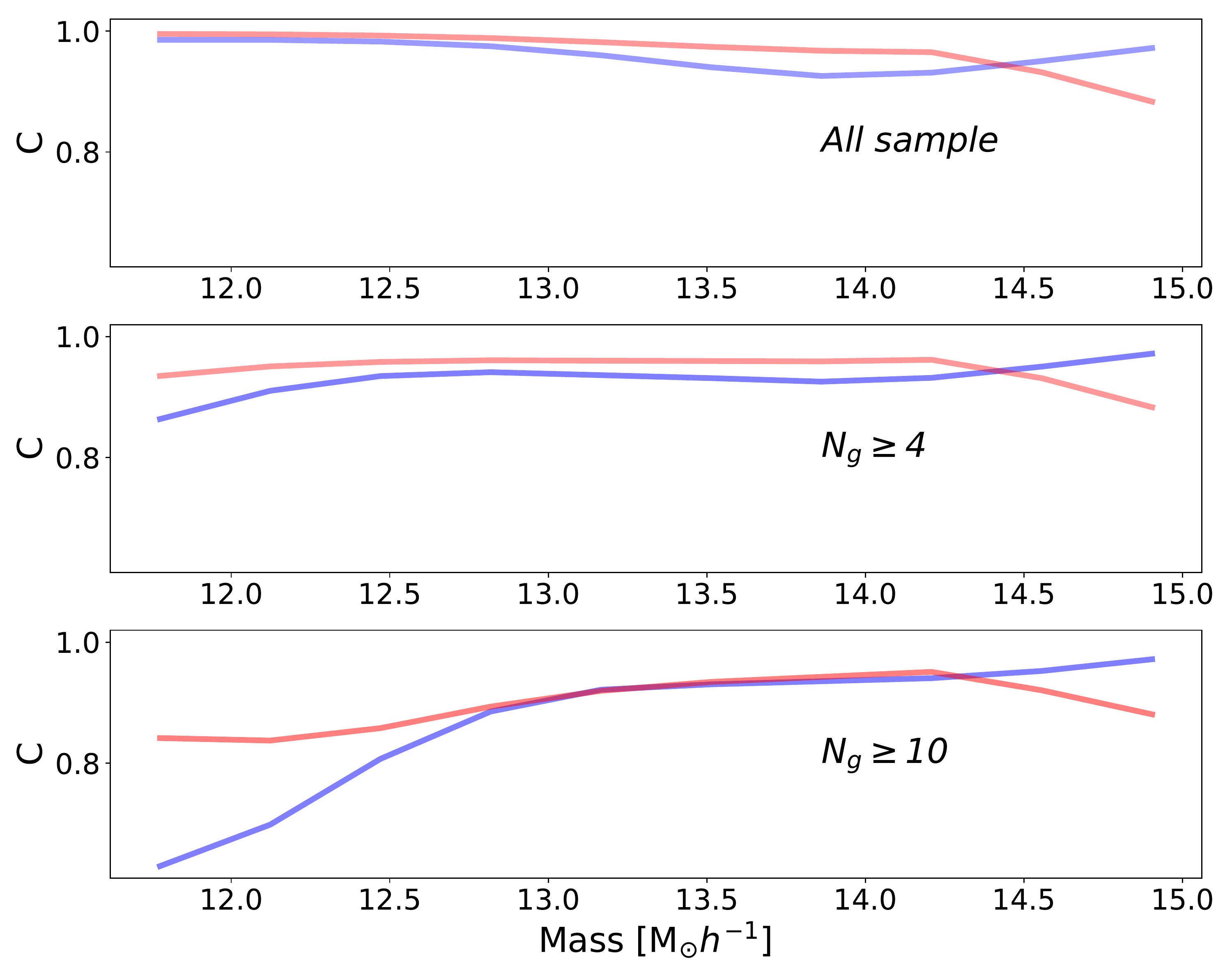}
   \includegraphics[width=0.49\textwidth]{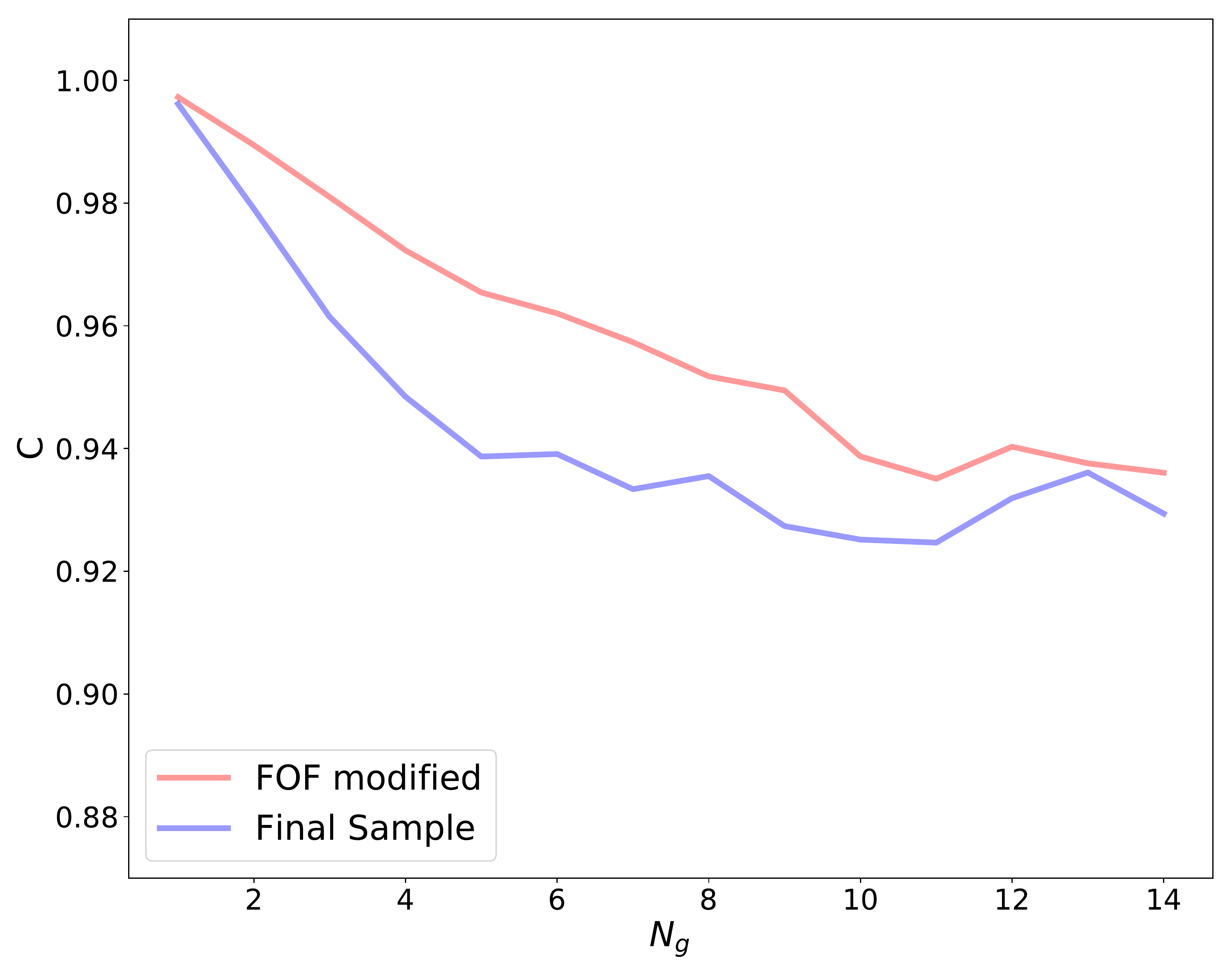}
      \caption{Halo completeness as a function of halo mass determined by abundance matching on luminosity for the FOF modified group catalogue (red) and Final Sample, combining FOF and halo-based methods (blue). Left panels show the results for all sample, groups with four or more members and groups with ten or more (top, middle, and bottom panels, respectively). The right panel presents halo completeness as a function of group multiplicity for the FOF modified group catalogue (red) and Final Sample (blue).}
         \label{Completeness}
   \end{figure*}

where $N_h$ is the total number of visible galaxies in the halo $h$ (i.e. the number of survivor galaxies of the halo once the apparent magnitude cut-off has been applied) and $ n_ {hg} $ is the number of visible galaxies in the halo $h$ that were also identified as members of the group $g$. As in the case of purity, if the halo shares galaxies with more than one identified group, we choose the group with the largest $n_ {hg}$. In figure \ref{Completeness} we show $ C $ following the same approach as in figure \ref{Purity}. In the left panels, we present $C$ as a function of mass and as can be seen, we obtain high completeness overall. However, the second step of the identification process tends to slightly deteriorate $C$ for low masses, particularly for those groups with high multiplicity. This is a natural consequence of the stringency imposed by the halo-based method, which particularly tends to rule out galaxies in the periphery of low-mass groups. On the other hand, the completeness of the most massive groups improves with the implementation of the entire process. This is because the density contrast, $ P_M (D, z) $, used to assign membership depends strongly on the mass. Similarly to the description of purity, in the right panel of Figure \ref{Purity} we also present $C$ as a function of multiplicity. The same arguments used to explain the behaviour of completeness as a function of mass can also be used to explain the behaviour of $C$ as a function of multiplicity shown in the right panel of Figure \ref{Completeness}, where there is a slight `worsening' of $C$  up to at least $N_g=14$ from where the plot becomes too noisy.

\subsection{Cumulative multiplicity function and mass comparison}

\begin{figure}
   \centering
   \includegraphics[width=0.5\textwidth]{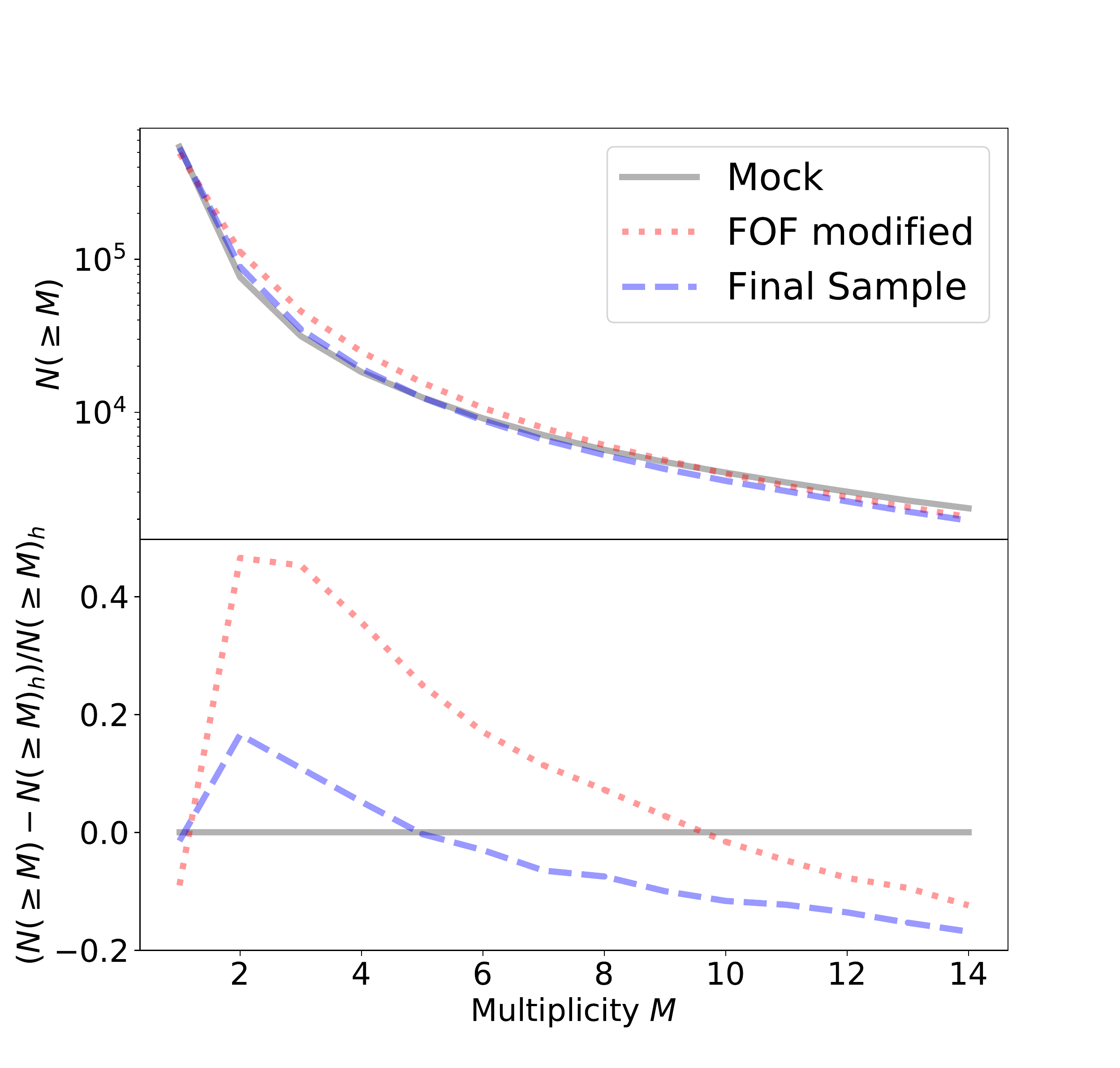}
      \caption{Cumulative multiplicity function, N($\geq$M), for halos in the mock sample (solid line), the FOF modified sample (dotted blue lines), and the Final Sample (dashed red lines). The bottom panel shows, for each multiplicity, the ratio between the difference of each sample with the mock sample and the total number of groups for the mock sample.}
         \label{Multiplicity}
   \end{figure}

Another test that we carried out on our group finder was to evaluate the cumulative multiplicity function, N($\geq$M), which provides information about the total number of groups that have more than a given number of members.
In Figure \ref{Multiplicity} we show the behaviour of this function for all our group catalogues: mock (solid grey lines), FOF modified (dotted red lines), and Final Sample (dashed blue lines). The upper panel presents the cumulative function for these samples, where it can be seen that the addition of the halo-based technique brings it closer to an ideal scenario, represented by groups taken from the simulation, $N(\geq M)_h$. To better demonstrate this behaviour we show in the bottom panel the relative difference between the identification and the result corresponding to the simulation: $(N(\geq M)_h-N(\geq M)_h)/N(\geq M)_h$. It is important to note that the main improvements to the implementation of the halo-based algorithm occur for groups with low multiplicities. This is a desired result since these groups have low reliability when they are identified with simple algorithms \citep[e.g.][]{Merchan2002, Merchan2005}.
For $M = 1$, this relationship shows us the difference between the total number of groups in the mock sample and those that result from each of the stages of our algorithm. Furthermore, it is observed that the number of groups identified by the complete algorithm that we propose is similar to the number of halos in the mock sample with at least one galaxy brighter than $M_r < -19.5$. However, for groups with high multiplicities, the same feature described in the previous section is manifested, namely the constraints imposed by the method on these groups tend to rule out the outermost galaxies.

\begin{figure}
   \centering
   \includegraphics[width=0.49\textwidth]{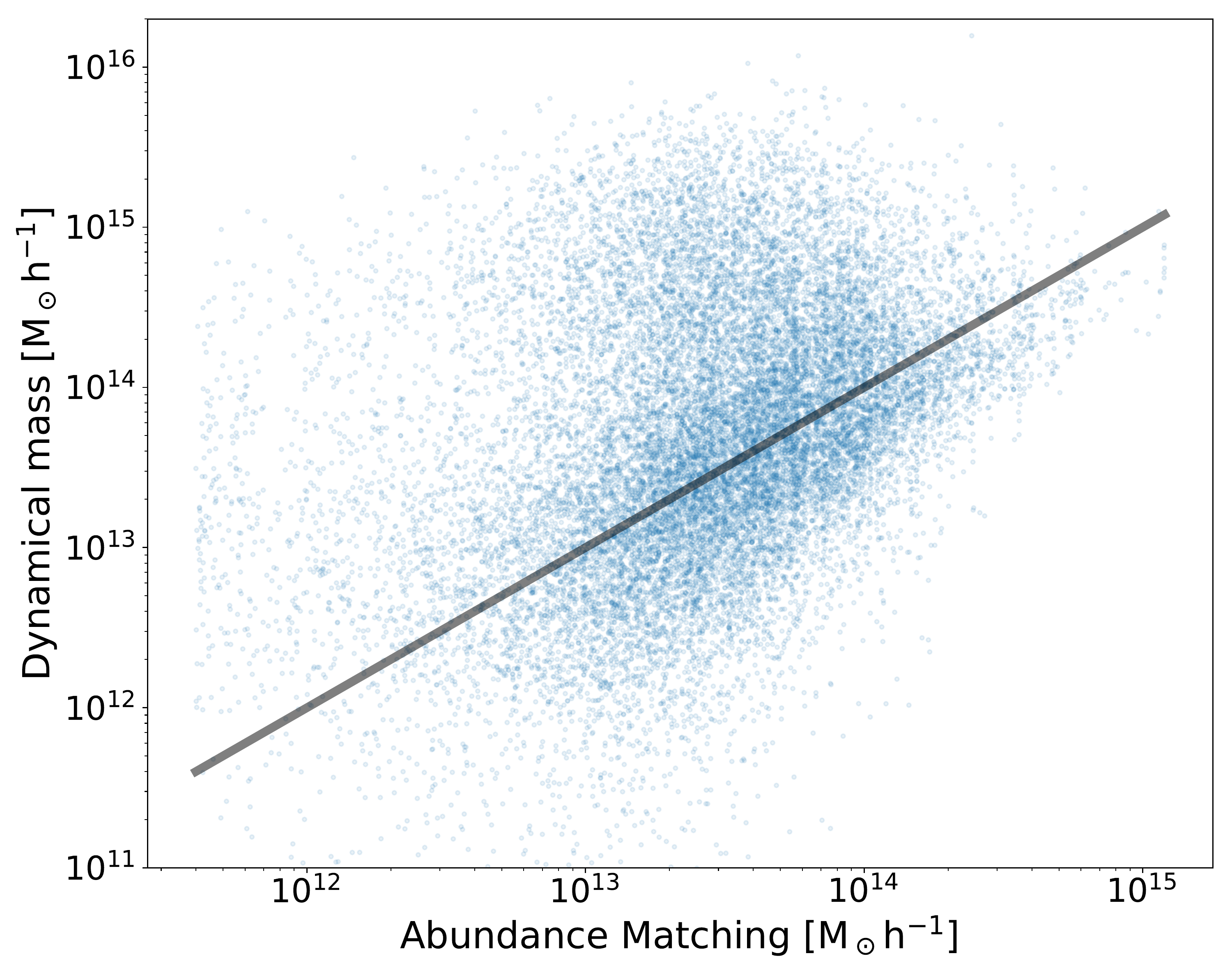}
      \caption{Comparison between the two techniques used to estimate the halo masses for groups with four or more members. The points show the mass obtained for each group following both methods and the grey line shows the one-to-one relationship to better visualise the dispersion.}
         \label{massmass}
   \end{figure}

The mass estimation of galaxy groups through their dynamic characteristics or brightness is a challenging topic given that the mass should be inferred indirectly from the limited information provided by the discrete number of galaxies available for each group. The output of our method allows us to easily estimate the masses following two different approaches. The simplest is the dynamic estimation using the positions and speeds of the member galaxies, which were calculated following \cite{Merchan2002}. This estimation is proportional to the product of the square of the velocity dispersion with the radius. On the other hand, the halo-based method naturally provides an estimation using the luminosity of the galaxy members to assign masses via the abundance-matching technique. A direct comparison of these two mass estimates is shown in the Figure \ref{massmass}. It should be taken into account that the dynamic mass is a statistic made from a small number of members and therefore the associated uncertainty is significant. This could be one of the main reasons for the observed dispersion in the figure; the influence of abundance matching uncertainties could be similar but its origins are more difficult to understand given the nature of this method.

   \begin{figure*}
   \centering
   \includegraphics[width=0.95\textwidth]{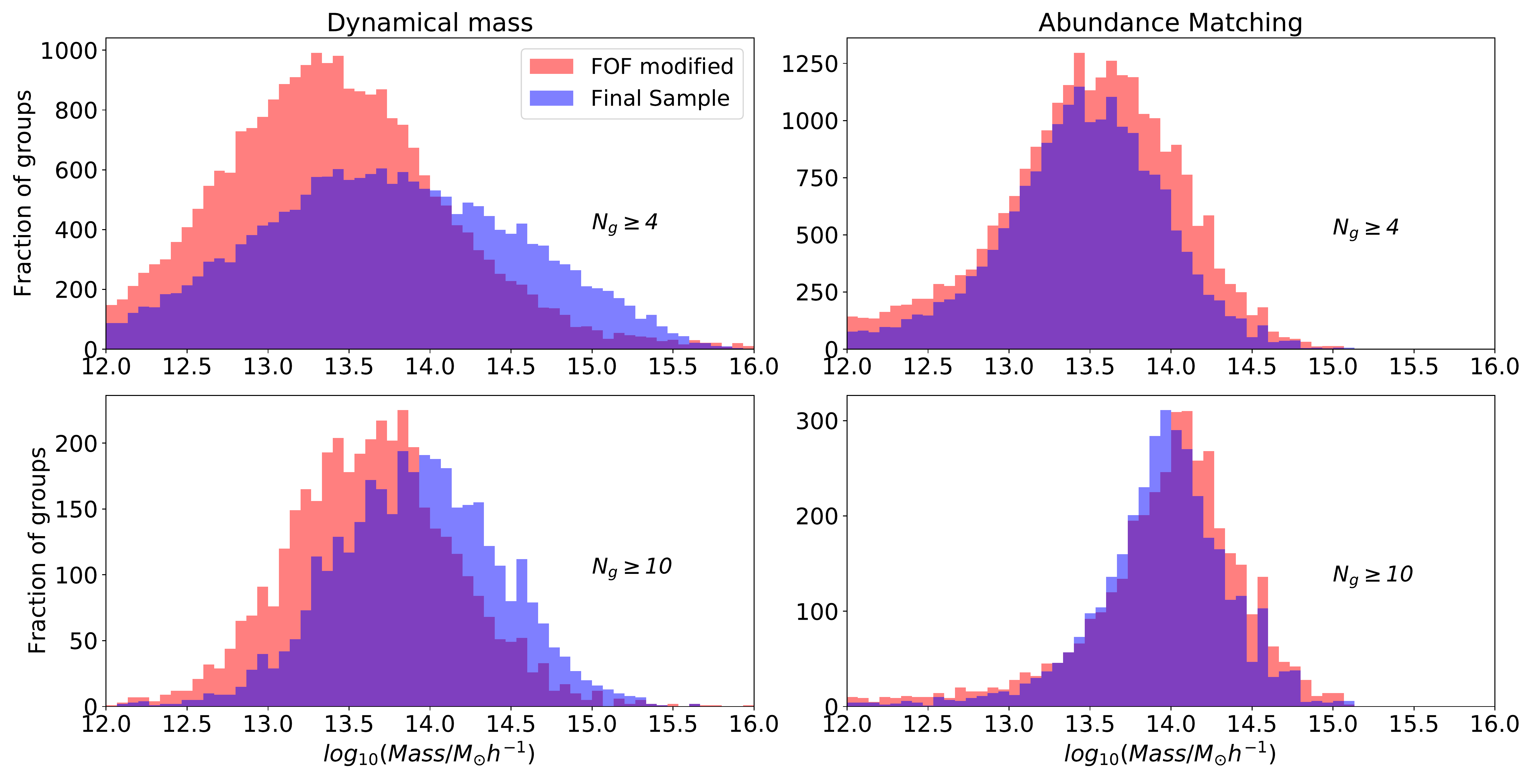}
      \caption{Mass Distributions. Left column: Result of dynamic mass estimates for  FOF modified (red) and Final Sample (blue) for groups with four or more members (upper panel) and ten or more (lower panel). Right column: Result of abundance matching mass for FOF modified (red) and Final Sample (blue) for groups with four or more members (upper panel) and ten or more (lower panel). }
         \label{mass}
   \end{figure*}

Now we compare the dynamical and abundance matching masses for the results of part I and II of our algorithm. Figure \ref{mass} shows results for both mass estimates: dynamical on the left and abundance matching on the right. We present the results for these samples because dynamic estimation with less than four members has a high uncertainty level, and we know that the mass of groups with ten or more galaxies is reliable for both estimations.  

If we  only pay attention to dynamical masses, we can observe that those obtained by part I (red) of the procedure are typically smaller than those of part II (blue). This can be explained by the same reasoning given earlier in this section, that is to say the halo-based method tends to reject external galaxies and keep the inner regions of groups, which have higher velocity dispersions. On the other hand, the virial radius estimation is dominated by internal galaxies, so when we take out the outer galaxies, it remains approximately constant. In addition, there is a large number of spurious groups with a low number of members that tend to be eliminated by the halo-based method. That is why the upper panel presents a different behaviour than the lower one, but both show that the mass increases. 

Looking now to the abundance matching masses in the right panels of Fig \ref{mass}, they seem to be more stable and show less variation between part I (red) and II (blue) of the algorithm. The removal of the outermost galaxies by part II of the procedure decreases the individual group luminosities and consequently a slight mass shift is produced in the mass distribution. This effect is similarly observed for groups of high and low multiplicity (upper and lower right panels of Fig \ref{mass}).

\subsection{Halo occupation distribution}

One of the main reasons for spending time and effort on developing an identifier capable of including groups of galaxies that traditional FOF algorithms cannot reliably detect ($N_g<4$) is its use in the study of the halo occupation distribution (HOD).

   \begin{figure*}
   \centering
      \includegraphics[width=0.95\textwidth]{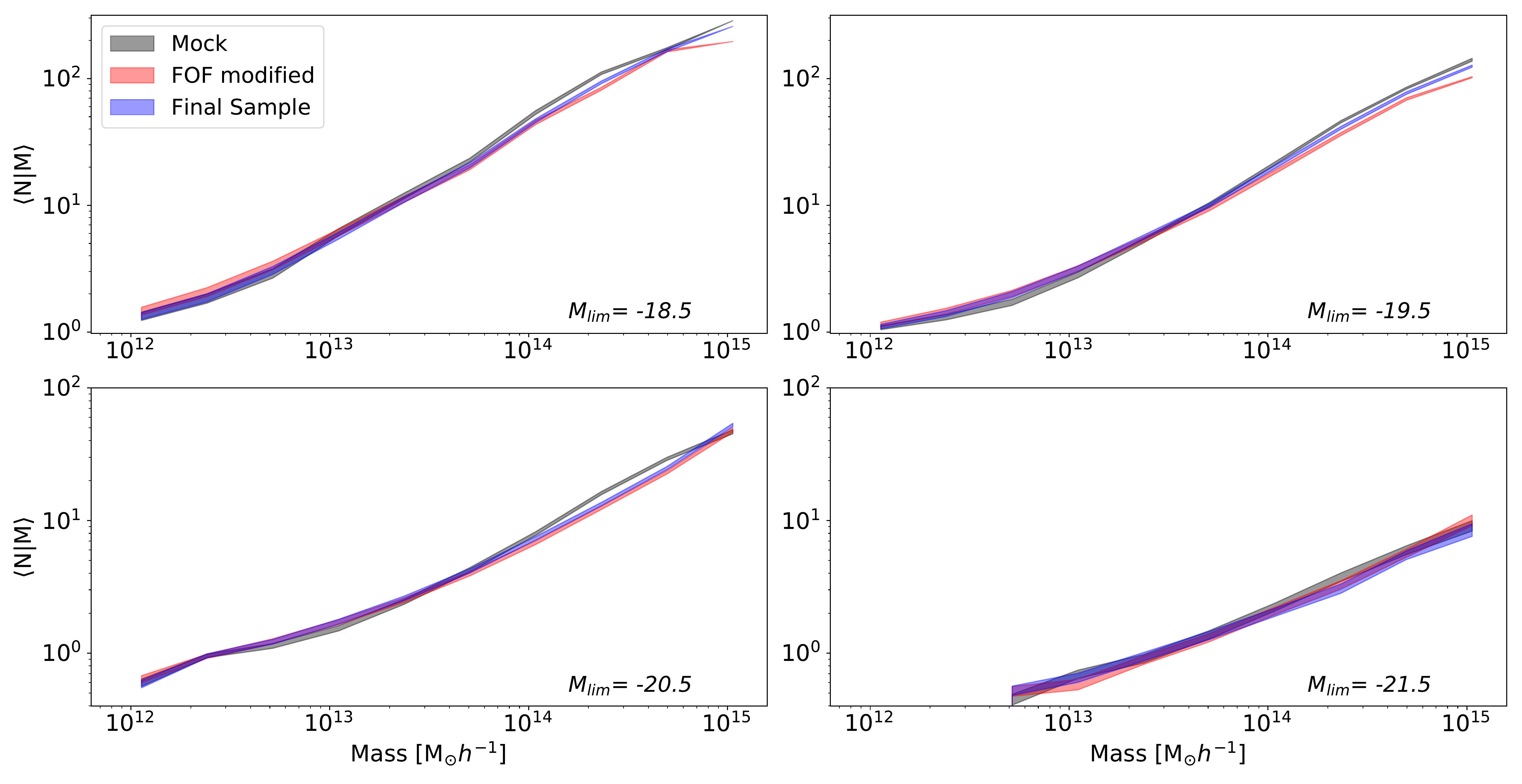}
      \caption{HOD for halos in the mock sample (grey), FOF modified (red) sample, and Final Sample (blue) for four absolute magnitude threshold ($M_r=$-18.5,-19.5,-20.5 and -21.5) as indicated in each panel.
      }
         \label{HOD}
   \end{figure*}

In Figure \ref{HOD} we compare the HOD for the catalogues presented in previous sections for four absolute magnitude thresholds ($M_{\rm lim}$). 
Since we need groups in the entire range of multiplicity, dynamical masses cannot be used, and therefore abundance matching masses are adopted for this purpose. 
As can be seen, the HOD taken from the FOF modified (red) and the Final Sample (blue) catalogues recover the corresponding to mock catalogue (grey). 
This result suggests that a suitable group definition and a reliable mass estimation is more important than the singularities of each of the parts of our algorithm.

\section{Application to the SDSS}
\label{sec:real}

Having a catalogue of reliable galaxy groups allows a wide range of scientific possibilities, ranging from the study of galaxy properties in dense media to large-scale structure studies, both in simulated catalogues and observational data. One of the most self-evident implementations is obtaining the SDSS groups. This photometric and spectroscopic survey  is the largest publicly available galaxy catalogue. Here, we specifically explore SDSS DR12. We take spectroscopic galaxies from the Legacy footprint area which covers more than 8400 $\rm{deg}^{2}$ in five optical bandpasses. After restricting the catalogue for our purposes (i.e. $\rm{z} <0.3$ and at $\rm{m_r}<17.77$), we hold more than $\sim 800 000$ objects.

To apply our algorithm to observational data, it is necessary to adopt a set of cosmological parameters and a theoretical mass function.
In our case, we adopt the WMAP3 cosmology \cite[$\Omega_m=0.238$, $\Omega_{\Lambda}=0.762$, $\Omega_{b}=0.042$, $n=0.951$, $h=H_0/(100 \rm{km s^{-1} Mpc^{-1}})=0.73$ and $\sigma_8=0.75$;][]{Spergel2007}, the halo mass function taken from \cite{Warren2006} and transfer function of \cite{Eisenstein1998}. It is important to note that our method is not sensitive to these parameters.

As a result of the entire process, a sample of 373252 groups with one or more members was obtained, a sample of 11941 with four or more, and a sample of  1666 with ten or more. For comparison, we perform the same procedure using in the mock sample, including the initial FOF modified catalogue. Both the Final sample and the FOF modified catalogues are available at \url{http://iate.oac.uncor.edu/alcance-publico/catalogos/}.

\begin{figure*}
   \centering
    \includegraphics[width=0.95\textwidth]{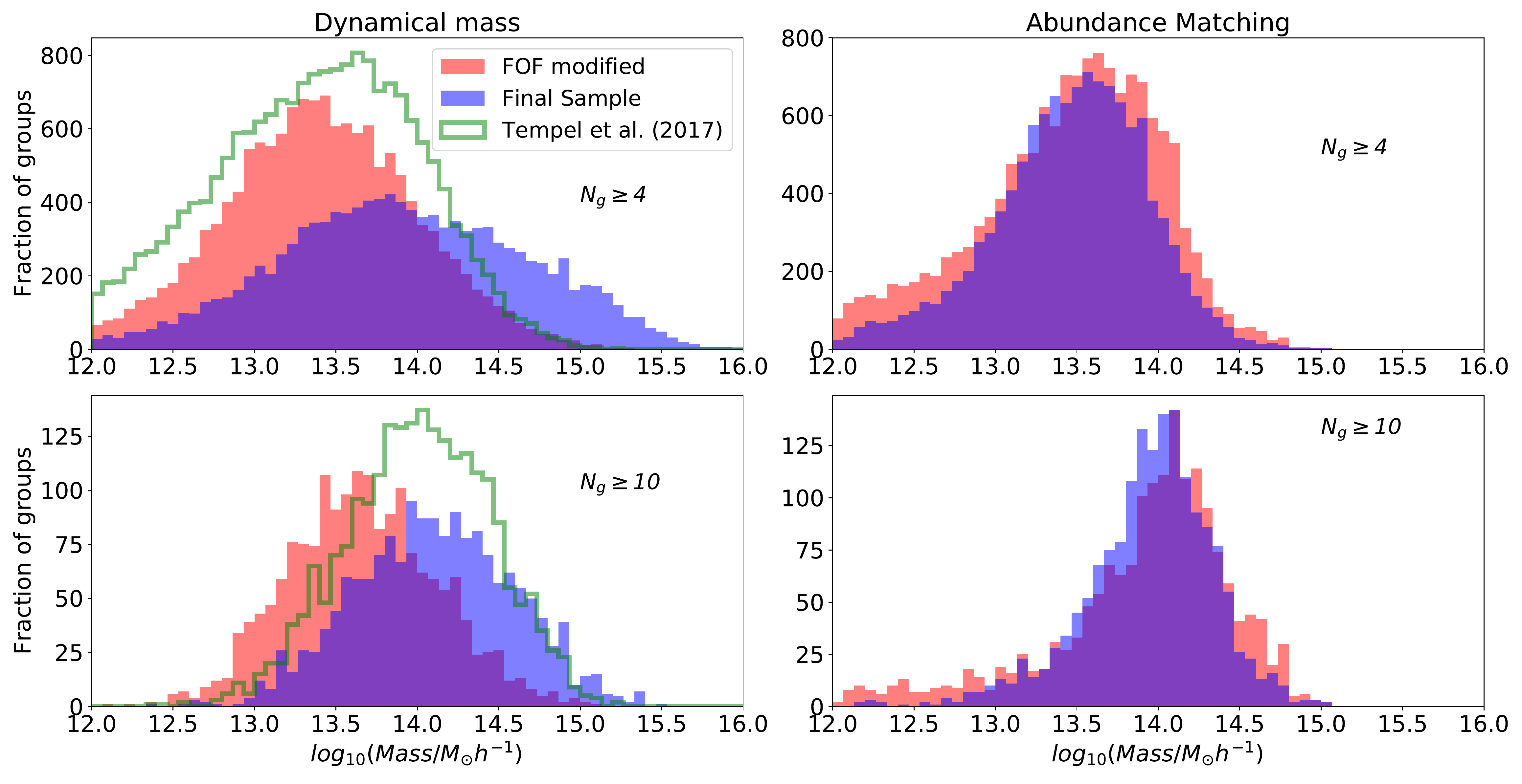}

      \caption{Mass distributions on SDSS DR12. The left column shows the result of dynamic mass estimates for  FOF modified (red), Final Sample (blue), and groups of \cite{Tempel2017} (green) for groups with four or more members (upper panel) and ten or more (lower panel). The right column shows the result of abundance matching mass for FOF modified (red) and Final Sample (blue) for groups with four or more members (upper panel) and ten or more (lower panel).}
         \label{Figmassreal}
   \end{figure*}

Following the same procedure as in the mock sample, in Figure \ref{Figmassreal} we present the mass distributions obtained for each catalogue using a dynamic and abundance matching mass estimations. 
As can be seen, the behaviour is very similar to that seen for the mock sample. In addition to the fact that the abundance matching method (right panel) allows estimation of the masses for groups with a multiplicity as low as one, it is less affected by the interlopers.

\begin{figure}
   \centering
    \includegraphics[width=0.49\textwidth]{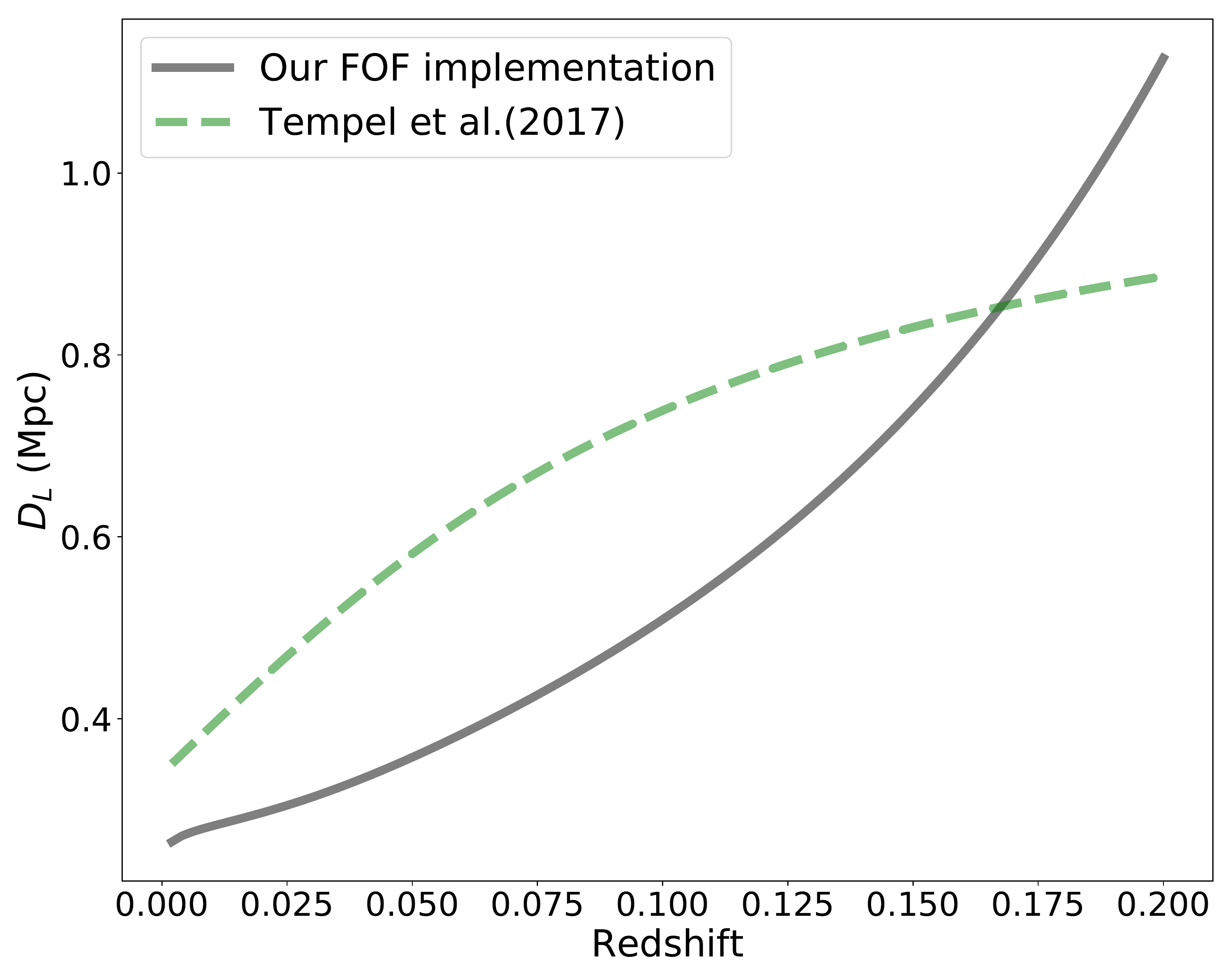}
     \caption{Comparison between the transverse linking length (D$_L$) in our FOF implementation (solid grey line) and those used by \cite{Tempel2017}: D$_L=0.34$Mpc$*[1+1.4*\texttt{arctan}(z/0.09)]$ (dashed green line).}
      \label{dl}
   \end{figure}

Taking advantage of the  availability of the \cite{Tempel2017} group catalogue, which was put together using the same data and following a logic similar to ours, we decided to compare the number of groups and the distribution of their masses with those obtained by us. The algorithm presented by these authors starts with a conventional FOF group identification improved through a refinement procedure that allows us to evaluate galaxy membership to higher-reliability groups. Although both our method and that of \cite{Tempel2017} use a FOF identification as a starting point, it should be noted that the transverse linking length (D$_L$) of the latter is considerably larger than ours in almost the entire redshift range, as can be seen in Fig \ref{dl}. Therefore, it is natural that they obtain a greater number of groups with two or more members (it must be taken into account that their catalogue does not have groups with one galaxy). In addition, we imposed to our initial FOF algorithm that groups have at least one bright galaxy, increasing the difference in the number of groups with the conventional FOF. However, one similarity of the two methods is the way in which  the dynamic mass is calculated and we expect the richest groups to be comparable. The number of groups with a multiplicity of $N\geq 4$ and $N\geq 10$ that \cite{Tempel2017} extract is 19775 and 2285, respectively, while our sample has 11941 and 1666, respectively. This can be seen in the dynamic mass distribution of Figure \ref{Figmassreal} (left panels). For the richest groups (ten or more galaxies, lower panel) the distribution of masses obtained with both methods shows a similar behaviour, that is, the refinements introduced to the original FOF seem to go in the same direction. The same does not happen for those groups that have four or more galaxies, where the groups of Temple et al. (green) are more similar to those of our FOF identification (red). This may be because our method uses masses obtained from the luminosity and therefore tends to discard low-luminosity groups, that is, those of low multiplicity.

\begin{figure*}
   \centering
   \includegraphics[width=0.95\textwidth]{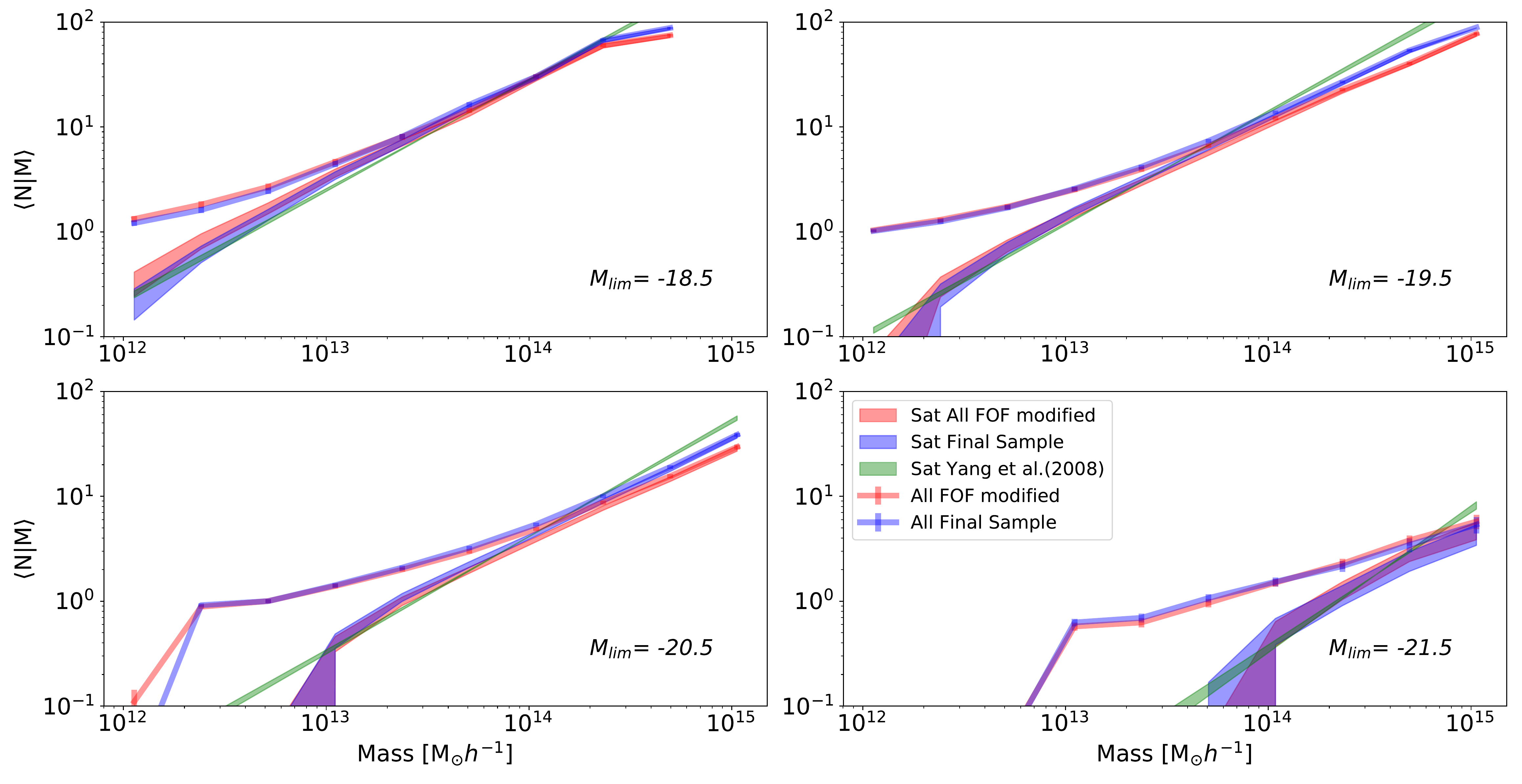}
      \caption{Halo occupation distribution for SDSS DR12 groups for FOF modified group sample (red lines) and Final Sample (blue lines) for four absolute magnitude thresholds ($M_r=$-18.5,-19.5,-20.5 and -21.5) as indicated in each panel. Green shaded areas correspond to the HOD of satellite galaxies obtained by \cite{Yang2008} in SDSS DR4, and to compare with our results shaded red and blue areas represent our HOD for satellites galaxies for the FOF modified and the Final Sample, respectively.}
         \label{Fighodreal}
   \end{figure*}

A simple way to compare our results with those of other authors is through the HOD.
In Figure \ref{Fighodreal} we show HOD measurements in our catalogues for four limits of absolute magnitude for the FOF modified group sample (red lines) and the  Final Sample (blue lines). Both are very similar to each other in all ranges of absolute magnitude studied (from $M_r=$-17 to -21.5), as can be seen in the four inner panels of the figure.
To compare with similar results, we include in this figure the occupation of the satellite galaxies described by \cite{Yang2008} for SDSS DR4 (green shaded area) together with the results corresponding to our groups (blue and red shaded areas for FOF modified and Final Sample, respectively).
As expected, given that the second part of our procedure is very similar to that of \cite{Yang2005}, the results of our entire procedure tend to resemble those of these latter authors. However, the improvements introduced by the halo-based method do not seem to be decisive for the HOD estimation.

\section{Summary and conclusions}
\label{sec:conc}
The goal of our research is to develop an algorithm that combines the advantages of the two most popular techniques in galaxy group identification: FOF and halo-based algorithms. Furthermore, as a simple application, we extracted and made publicly available a catalogue of galaxy groups taken from the SDSS DR12, where, in addition to the main properties of the groups, member galaxies are listed with the corresponding identification label from the original catalogue.

We asses our algorithm using a mock catalogue of galaxies, obtaining excellent results in terms of purity and completeness. Given that our method provides two output catalogues (FOF modified and Final Sample) we were able to compare the results corresponding to each one of them.
The good performance of the FOF modified algorithm allowed high levels of reliability to be obtained with an average purity of greater than 0.7 together with a completeness of greater than 0.9. Nevertheless, the entire process improves purity at the price of a small loss of completeness. Another way to compare the performance of both catalogues is to analyse the multiplicity, where we find that the number of groups with a given number of members is significantly closer to that of the mock sample in the case of the Final Sample.

An expected result is that both the dynamic mass and that obtained with the abundance matching method are similar when it comes to high multiplicities. However, for low multiplicities, the abundance matching masses behave better since they are not influenced by the uncertainties introduced into the dynamic masses by the low-number statistical error in the velocity dispersion and the virial radius.
In addition, dynamic masses are more influenced by interlopers since the luminosity used in the abundance matching method is dominated by brighter galaxies.

To conclude with the evaluation of the method, we compare the HODs obtained from our catalogues with those corresponding to the mock sample. We were able to reproduce their behaviour in a wide range of masses and magnitudes for both catalogues. Additionally, we find that HOD does not depend strongly on the identification method used.

An important added value of this work is that we apply this algorithm to the SDSS DR12, obtaining reliable group catalogues that we make publicly available. Both the FOF and the Final Sample catalogues include the main properties of the groups such as abundance matching mass, dynamic mass (for those with $N_g\geq$4), virial radius, and velocity dispersion. Separately, member galaxies from each group are also included. 

The proposed algorithm can be applied to any catalogue of galaxies that lists positions, magnitudes, and redshifts. Particularly, it could be easily used to extract galaxy systems from the next generation of redshift surveys. 

\begin{acknowledgements}

We thank the referee, Elmo Tempel, for insightful comments.
This work was also partially supported by Agencia Nacional de Promoci\'on Cient\'ifica y Tecno\'oogica (PICT 2015-3098), the Consejo Nacional de Investigaciones Cient\'{\i}ficas y T\'ecnicas (CONICET, Argentina) and the Secretar\'{\i}a de Ciencia y Tecnolog\'{\i}a de la Universidad Nacional de C\'ordoba (SeCyT-UNC, Argentina). 

\end{acknowledgements}

\bibliographystyle{aa} 
\bibliography{Yourfile} 


\appendix

\section{Halo properties: Mass, radius, and velocity.}
\label{appendixx}

If we define the dark matter halos as bounded spherical structures of an overdensity ($\Delta$), the virial theorem can be used to estimate their mass (M$_\Delta$) and, from this, obtain the radius (r$_\Delta$) and velocity dispersion ($\sigma _\Delta$) associated with each halo. The mass can be written according to its density as: 
\begin{equation}
M_{\Delta}=4 \pi \int _0 ^{r_\Delta} r^2 \rho (r) dr
.\end{equation}
The above expression can be written based on $\Delta$ and critical density of the universe ($\rho _c$):
\begin{equation}
M_{\Delta}=\frac{4}{3} \pi \,  \, \rho _c  \, r ^3_\Delta \, \Delta
.\end{equation}
If we consider that $\rho _c =\frac{3\, H^2(z)}{8\, \pi \, G}$, we can rewrite the last expression as:
\begin{equation}
M_{\Delta}=\frac{ \, H^2(z) \, r ^3_\Delta \, \Delta}{2 \, G} 
.\end{equation}
From this last expression, $r_\Delta$ can be expressed as: 
\begin{equation}
r_{\Delta}=\left(\frac{2 \, M_{\Delta} \, G }{H^2(z) \, \Delta} \right) ^\frac{1}{3} 
.\end{equation}
Once this radius is calculated, the circular velocity ($\sigma _\Delta$ ) to $r_{\Delta}$ can be obtained:
\begin{equation}
\sigma_{\Delta}=\sqrt{\frac{M_{\Delta} \, G }{r_{\Delta}}}
.\end{equation}
Subsequently, the line-of-sight velocity dispersion ($\sigma$) is reduced to:
\begin{equation}
\sigma =\frac{\sigma_{\Delta}}{\sqrt{3}}
.\end{equation}

\end{document}